\documentclass[aps, pra, twocolumn, floatfix, nofootinbib, showpacs]{revtex4}
\usepackage{graphicx}
\usepackage{amsmath}
\begin{document}
\title{The effect of hyperfine interactions on ultracold molecular collisions:\\
NH$(^3\Sigma^-)$ with Mg($^1$S) in magnetic fields}

\author{Maykel L.~Gonz\'alez-Mart\'{\i}nez}
\affiliation{Department of Chemistry, Durham University, Durham DH1~3LE, United
Kingdom}

\author{Jeremy M.~Hutson}
\affiliation{Department of Chemistry, Durham University, Durham DH1~3LE, United
Kingdom}

\date{\today}
\begin{abstract}
\noindent We investigate the effect of hyperfine interactions on
ultracold molecular collisions in magnetic fields, using
$^{24}$Mg($^1$S) + $^{14}$NH($^3\Sigma^-$) as a prototype system. We
explore the energy and magnetic field dependence of the cross sections,
comparing the results with previous calculations that neglected
hyperfine interactions (Phys.~Rev.~Lett.~\textbf{103}, 183201 (2009)).
The main effect of hyperfine interactions for spin relaxation cross
sections is that the kinetic energy release of the dominant outgoing
channels does not reduce to zero at low fields. This results in reduced
centrifugal suppression on the cross sections and increased inelastic
cross sections at low energy and low field. We also analyze
state-to-state cross sections, for various initial states, and show
that hyperfine interactions introduce additional mechanisms for spin
relaxation. In particular, there are hyperfine-mediated collisions to
outgoing channels that are not centrifugally suppressed. However, for
Mg+NH these unsuppressed channels make only small contributions to the
total cross sections. We consider the implications of our results for
sympathetic cooling of NH by Mg and conclude that the ratio of elastic
to inelastic cross sections remains high enough for sympathetic cooling
to proceed.
\end{abstract}


\maketitle
\section{Introduction}
\label{sec:introduction} High-density samples of cold ($T < 1$~K) and
ultracold ($T < 1$~$\mu$K) molecules are likely to have important
applications in fields including quantum information science,
high-precision spectroscopy and quantum-controlled chemistry
\cite{ldcarr:09b}. For nearly three decades, ultracold atoms have been
produced using laser Doppler cooling and evaporative cooling.  At
temperatures between 1~nK and 1~$\mu$K, atoms enter a fully quantal
regime with novel properties, forming Bose-Einstein condensates
\cite{mhanderson:95,ccbradley:95,kbdavis:95} and Fermi-degenerate gases
\cite{bdemarco:99,agtruscott:01}.  However, cooling molecular samples
to the temperatures required for quantum degeneracy is much more
difficult than for atoms, because molecules have a much richer internal
structure.

Molecular cooling techniques can be classified as either \emph{direct}
or \emph{indirect}.  In direct methods, molecules are cooled from
relatively high temperatures by techniques such as buffer-gas cooling
\cite{jdweinstein:98a}, Stark deceleration \cite{hlbethlem:03} or
Zeeman deceleration \cite{nvanhaecke:07,sdhogan:07,enarevicius:07}.  In
addition, laser Doppler cooling has very recently been demonstrated for
SrF \cite{esshuman:10}, although only a very limited number of
molecular species are likely to be amenable to this technique. Indirect
methods, by contrast, form ultracold molecules in gases of previously
cooled atoms by magnetoassociation \cite{jmhutson:06,tkohler:06} or
photoassociation \cite{kmjones:06,jmhutson:06}.  In the last few years,
indirect methods in association with stimulated Raman adiabatic passage
(STIRAP) \cite{kbergmann:98} have succeeded in producing ground-state
molecules at temperatures below 1~$\mu$K
\cite{k-kni:08,jdeiglmayr:08,jgdanzl:10}. Indirect methods are however
restricted to the relatively few atomic species that can be
laser-cooled.

Direct methods can in principle be used for a wide variety of
molecules, but the lowest temperatures achieved so far are in the
millikelvin regime.  A \emph{second-stage} cooling technique is needed
to cool molecules down to the microkelvin range, from which evaporative
cooling could reach the region of quantum degeneracy. \emph{Sympathetic
cooling} is a promising second-stage cooling method, which relies on
thermalization of the molecular species of interest by collisions with
ultracold atoms.  Sympathetic cooling has been used for almost three
decades to cool molecular ions to temperatures around 100~mK and below
\cite[ch.~18]{rvkrems:book09}, and has more recently also been used for
neutral atoms \cite{cjmyatt:97,gmodugno:01}.  Its extension to neutral
molecules was first proposed in 2004 by Sold\'an and Hutson
\cite{psoldan:04}.  Since then, accurate quantum scattering
calculations have suggested that it should work for a few diatomic
\cite{aogwallis:09b,pszuchowski:11a} and polyatomic molecular species
\cite{pszuchowski:09a,tvtscherbul:11a} in combination with appropriate
ultracold atoms. Although experimental validation is still needed,
there is great hope that sympathetic cooling will eventually become a
`routine' second-stage cooling technique, capable of producing large
molecular samples in the microkelvin regime, where most of the new
interesting applications would become possible \cite{ldcarr:09b}.

Cold atoms and molecules are usually held in traps formed with
electric, magnetic or optical fields. By far the most experimentally
accessible traps are those formed by static electric
\cite{hlbethlem:00b} and magnetic fields \cite{jdweinstein:98a}. The
main limitation of such traps is that they can hold only those
molecules whose energy \emph{increases} with applied field, \emph{i.e.}
those in low-field-seeking states. However, there are always untrapped
high-field-seeking states that lie energetically below the
low-field-seeking states. Inelastic (deexcitation) collisions that
transfer molecules to high-field seeking states therefore lead to trap
loss. Elastic collisions, by contrast, are required for thermalization.
Thus, the success of both sympathetic and evaporative cooling depends
on a favorable ratio of elastic to inelastic cross sections, of at
least about 100 \cite{rdecarvalho:99}.

In the (ultra)cold regime, hyperfine interactions are larger than or
comparable to the kinetic energies involved. Hyperfine couplings also
provide both additional relaxation mechanisms and new ways to control
atomic and molecular collisions. However, only a few calculations have
so far considered the consequences of hyperfine interactions on cold
molecular collisions. Bohn and coworkers considered the scattering of
polar $^{16}$OH($^2\Pi_{3/2}$) molecules in static electric
\cite{avavdeenkov:02} and magnetic fields \cite{cticknor:05a}. Lara
\emph{et al.}\ \cite{mlara:06,mlara:07} studied the field-free
scattering of Rb($^2$S) and OH($^2\Pi_{3/2}$). Tscherbul \emph{et
al.}~studied (ultra)cold collisions of YbF($^2\Sigma$) molecules with
He in external electric and magnetic fields \cite{tvtscherbul:07}. They
found that simultaneous electron and nuclear spin relaxation for
collisions in the ground rotational state can occur through couplings
via excited rotational states and hyperfine terms.

In the present paper, we explore the effects of hyperfine interactions
on the ultracold scattering of NH with Mg in the presence of a magnetic
field. This extends the study in \cite{tvtscherbul:07} in various ways:
(1) NH is a $^3\Sigma$ instead of a $^2\Sigma$ molecule; (2) NH has two
nuclei with non-zero spin, leading to more complicated hyperfine
structure than for YbF; (3) we consider hyperfine states other than
spin-stretched states. Our results are compared with those of a
previous study of Mg-NH neglecting hyperfine terms
\cite{aogwallis:09b}, and allow us to make some general conclusions on
how hyperfine interactions can affect the prospects of sympathetic
cooling.

\section{Theory}
\label{sec:theory} We consider the scattering of
$^{14}$NH($^3\Sigma^-$) molecules with $^{24}$Mg($^1$S) atoms in the
presence of an external magnetic field $B$, whose direction defines the
space-fixed $Z$-axis.  The system is described using Jacobi
vectors/coordinates: the vector $\boldsymbol{r}$ runs from N to H while
$\boldsymbol{R}$ runs from the center of mass of NH to Mg. The angle
between $\boldsymbol{r}$ and $\boldsymbol{R}$ is $\theta$. The
collision is studied by solving the time-independent Schr\"odinger
equation for the scattering wave function $\Psi$ at energy $E$,
$\hat{\mathcal{H}}\Psi = E\Psi$, where $\hat{\mathcal{H}}$ is the
Hamiltonian for the colliding pair.

\subsection{Effective Hamiltonian}
\label{sec2:Heff} By convention, lower-case symbols are used for
quantum numbers of the individual monomers, and capital letters are
used for quantum numbers of the collision complex as a whole. Where
necessary, the subscripts 1 and 2 refer to Mg and NH respectively. The
diatom is considered to be a rigid rotor in its ground vibrational
state.  The effective Hamiltonian can be written
\begin{equation}
 \hat{\mathcal{H}} = - \frac{\hbar^2}{2\mu} R^{-1} \frac{d^2}{dR^2} R
                     + \frac{\hbar^2\hat{L}^2}{2\mu R^2}
                     + \hat{\mathcal{H}}_\mathrm{mon}
                     + \hat{\mathcal{H}}_{12},
 \label{eq:Heff}
\end{equation}
where $\hat{L}$ is the space-fixed operator for the end-over-end
rotation and $\mu$ is the reduced mass of the complex. In general,
$\hat{\mathcal{H}}_\mathrm{mon}$ contains terms describing both
isolated monomers, $\hat{\mathcal{H}}_\mathrm{mon} =
\hat{\mathcal{H}}_1 + \hat{\mathcal{H}}_2$.  However, ground-state
$^{24}$Mg has both zero electron spin and zero nuclear spin, so that it
contributes only a constant energy and it is convenient to set
$\hat{\mathcal{H}}_1 = 0$. Finally, $\hat{\mathcal{H}}_{12}$ includes
all interactions between the monomers, which in the present case
reduces to the potential energy surface
$\hat{V}(\boldsymbol{u}_r,\boldsymbol{R})$, conveniently expanded in
Legendre polynomials
\begin{eqnarray}
  \hat{V}(\boldsymbol{u}_r,\boldsymbol{R}) = \sum^{k_\mathrm{max}}_{k=0}
   V_k(r_\mathrm{eq},R) P_k(\cos{\theta}).
 \label{eq:VinLP}
\end{eqnarray}
Here $\boldsymbol{u}_r$ denotes a unit vector in the direction of
$\boldsymbol{r}$, $r_\mathrm{eq}$ is the equilibrium distance of NH in
its ground vibrational state and $V_k$ are the radial strength
functions.  The potential for the Mg-NH system was reported in
Ref.~\cite{psoldan:09}.

The Hamiltonian for an isolated NH($^3\Sigma^-$) molecule can be
written $\hat{\mathcal{H}}_2 = \hat{\mathcal{H}}_\mathrm{rot} +
\hat{\mathcal{H}}_\mathrm{sn} + \hat{\mathcal{H}}_\mathrm{ss} +
\hat{\mathcal{H}}_\mathrm{hf} + \hat{\mathcal{H}}_\mathrm{Z}$.  The
different terms will be discussed below and correspond, respectively,
to the rotational, electron spin-rotation, electron spin-spin,
hyperfine and Zeeman interactions.

If centrifugal distortion and all other higher-order corrections are
neglected, the Hamiltonian for the mechanical rotation of NH is
$\hat{\mathcal{H}}_\mathrm{rot} = b_\mathrm{NH} \hat{n}^2$, where
$b_\mathrm{NH}$ is the rotational constant (with dimensions of energy)
and $\hat{n}$ is the operator for the rotational angular momentum. The
electron spin-rotation term, arising from the interaction between the
magnetic moment associated with the composite electronic spin of NH,
$\hat{s}$, and the magnetic field induced by its rotation, can be
written $\hat{\mathcal{H}}_\mathrm{sn} = \gamma \hat{n} \cdot \hat{s}$,
where $\gamma$ is the spin-rotation constant.

The direct dipolar interaction between the unpaired electrons in
NH($^3\Sigma^-$) may be written \cite{jmbrown:book03}
\begin{equation}
 \hat{\mathcal{H}}_\mathrm{ss} \approx 2\lambda_\mathrm{ss}
  \mathrm{T}^2_{q=0}(\hat{s},\hat{s})
  \mathrm{T}^2_{q=0}(\boldsymbol{u}_r,\boldsymbol{u}_r),
 \label{eq:Hss_approx}
\end{equation}
where $\lambda_\mathrm{ss}$ is the electron spin-spin constant and
$\mathrm{T}^k$ represents a spherical tensor of rank $k$, with
$q$-component $\mathrm{T}^k_q$.

Since both the $^{14}$N and $^1$H nuclei have non-zero nuclear spin,
$i_\mathrm{N} = 1$ and $i_\mathrm{H} = 1/2$, the hyperfine Hamiltonian
can be written
\begin{equation}
\hat{\mathcal{H}}_\mathrm{hf} =
\hat{\mathcal{H}}_\mathrm{si,N} + \hat{\mathcal{H}}_\mathrm{F,N} +
\hat{\mathcal{H}}_\mathrm{in,N} + \hat{\mathcal{H}}_\mathrm{Q,N} +
\hat{\mathcal{H}}_\mathrm{si,H} + \hat{\mathcal{H}}_\mathrm{F,H} +
\hat{\mathcal{H}}_\mathrm{in,H}.
\end{equation}
Here, $\hat{\mathcal{H}}_\mathrm{si}$ represents the direct dipolar
interaction between the magnetic moments associated with a given
nuclear spin $\hat{\imath}$ and the composite electron spin of the
open-shell $\Sigma$ molecule and can be written \cite{jmbrown:book03}
\begin{eqnarray}
 \hat{\mathcal{H}}_\mathrm{si} &=& - \sqrt{10} g_S \mu_{\rm B} g_i \mu_{\rm N} (\mu_0/4\pi)
  \mathrm{T}^1(\hat{s},\boldsymbol{C}^2) \cdot \mathrm{T}^1(\hat{\imath})
  \nonumber\\
  &\approx& \sqrt{6} t_0 \mathrm{T}^2_{q=0}(\hat{s},\hat{\imath}),
 \label{eq:Hsi_approx}
\end{eqnarray}
where $g_S$ and $g_i$ are the electron and nuclear $g$-factors,
$\mu_{\rm B}$ and $\mu_{\rm N}$ are the Bohr and nuclear magnetons, and
$\mu_0$ is the magnetic permeability of free space.  The axial dipolar
interaction parameter $t_0$ is related to the widely-used constant $c$
defined by Frosch and Foley \cite{rafrosch:52} as $t_0 = c/3$.

The Fermi (or Fermi-Breit) contact interaction
$\hat{\mathcal{H}}_\mathrm{F}$ occurs whenever there is a non-zero
electron-spin density at a nucleus with non-zero spin $\hat{\imath}$.
It may be written $\hat{\mathcal{H}}_\mathrm{F} = b_\mathrm{F} \hat{s}
\cdot \hat{\imath}$, where $b_\mathrm{F}$ is a coupling constant given
by $b_\mathrm{F} = (2/3) g_S \mu_{\rm B} g_i \mu_{\rm N} |\psi(0)|^2$,
with $|\psi(0)|^2$ the spin density. The coupling constant can also be
written in terms of Frosch and Foley's $b$ and $c$ parameters as
$b_\mathrm{F} = b + (1/3)c$.

Since $i_\mathrm{N} > 1/2$, the interaction between the nuclear
electric quadrupole moment $\boldsymbol{Q}$ and the electric field
gradient at the $^{14}$N nucleus $\nabla\boldsymbol{E}$ must be
included.  In general, this is \cite{jmbrown:book03}
\begin{equation}
 \hat{\mathcal{H}}_\mathrm{Q} = -e \mathrm{T}^2(\nabla\boldsymbol{E}) \cdot
  \mathrm{T }^2(\boldsymbol{Q}),
 \label{eq:HQ}
\end{equation}
and for a diatomic molecule reduces to
\begin{equation}
 \hat{\mathcal{H}}_\mathrm{Q} = \frac{eq_0Q}{4i(2i-1)} \sqrt{6}
  \mathrm{T}^2_{q=0}(\hat{\imath},\hat{\imath}),
 \label{eq:HQ_approx}
\end{equation}
where $q_0$ is the electric field gradient.

The nuclear spin-rotation terms $\hat{\mathcal{H}}_\mathrm{in}$ are the
nuclear counterpart of the $\hat{\mathcal{H}}_\mathrm{sn}$ interaction
discussed above, and are written $\hat{\mathcal{H}}_\mathrm{in} =
C_\mathrm{I} \hat{n} \cdot \hat{\imath}$, where $C_\mathrm{I}$ is the
corresponding nuclear spin-rotation constant.

Finally, if only electron and nuclear spin Zeeman terms are taken into
account,
\begin{equation}
 \hat{\mathcal{H}}_\mathrm{Z} = g_S \mu_{\rm B} \hat{s}_z \boldsymbol{B}
  - \mu_{\rm N} \sum_\mathrm{X = N,\, H} g_\mathrm{X}
   \hat{\imath}_{\mathrm{X}z} \boldsymbol{B} (1-\sigma_\mathrm{X}).
 \label{eq:HZ}
\end{equation}
The nuclear shielding factors $\sigma_\mathrm{X}$ are extremely small
and are neglected in the present work.

The molecular constants used for the $^{14}$NH($^3\Sigma^-$) radical
are listed in Table~\ref{tab:NHparameters}.

\begin{table}[b!]
 \caption{Molecular parameters for $^{14}$NH($^3\Sigma^-$, $v = 0$).
 \label{tab:NHparameters}}
 \begin{ruledtabular}
  \begin{tabular}{crc}
   Parameter                                & Value     & Reference \\ \hline
   $b_\mathrm{NH}$/cm$^{-1}$                &  16.343   & \cite{mmizushima:book75} \\
   $\gamma$/cm$^{-1}$                       & $-0.055$  & \cite{mmizushima:book75} \\
   $\lambda_\mathrm{ss}$/cm$^{-1}$          &   0.92    & \cite{mmizushima:book75} \\
   $g_\mathrm{N}$                           &   0.40376 & \cite{IUPAC:2007} \\
   $b_\mathrm{F,N}$/MHz                     &  18.83    & \cite{jflores-mijangos:04} \\
   $c_\mathrm{N} (= 3t_{0,\mathrm{N}})$/MHz &$-67.922$  & \cite{jflores-mijangos:04} \\
   $(eq_0Q)_\mathrm{N}$/MHz                 & $-2.883$  & \cite{jflores-mijangos:04} \\
   $C_\mathrm{I,N}$/MHz                     &   0.1455  & \cite{jflores-mijangos:04} \\
   $g_\mathrm{H}$                           &   5.58568 & \cite{IUPAC:2007} \\
   $b_\mathrm{F,H}$/MHz                     &$-66.131$  & \cite{jflores-mijangos:04} \\
   $c_\mathrm{H} (= 3t_{0,\mathrm{H}})$/MHz &  90.291   & \cite{jflores-mijangos:04} \\
   $C_\mathrm{I,H}$/MHz                     & $-0.061$  & \cite{jflores-mijangos:04}
  \end{tabular}
 \end{ruledtabular}
\end{table}

\subsection{Coupled-channel equations}
\label{sec2:coupled-eqs} We solve the scattering problem using the
coupled-channel method.  First, the total wave function is expanded in
a set of $N$ conveniently chosen basis functions $\left|a\right>$,
\begin{equation}
\Psi(R,\xi) = R^{-1} \sum_{a} \chi_a(R) \left|a\right>.
\end{equation}
Here, $\xi$ is a collective variable including all coordinates except
$R$, and $a$ is the set of quantum numbers that label our basis
functions.  Each different combination of quantum numbers $a$ is said
to define a \emph{channel}. A set of coupled differential equations for
the \emph{channel functions} $\chi_a(R)$ is then obtained by
substituting $\Psi(R,\xi)$ into the time-independent Schr\"odinger
equation,
\begin{equation}
 \frac{d^2 \chi_a}{dR^2}
              = \sum_{a'} \left(W_{aa'} - \epsilon\delta_{aa'}\right) \chi_{a'},
 \label{eq:coupled-eqs}
\end{equation}
where $\delta_{ij}$ is the Kronecker delta, $\epsilon = 2\mu E/\hbar^2$
is a scaled energy, and
\begin{equation}
 W_{aa'}(R) = \frac{2\mu}{\hbar^2}
              \left<a\right| \left[ \hat{\mathcal{H}}_\mathrm{mon}
              + \hat{\mathcal{H}}_{12}
              + \frac{\hbar^2\hat{L}^2}{2\mu R^2} \right] \left|a'\right>.
 \label{eq:waa'}
\end{equation}
The coupled equations \eqref{eq:coupled-eqs} are solved by propagating
a complete set of independent solution vectors from $R_\mathrm{min}$,
deep in the inner classically forbidden region, to $R_\mathrm{max}$,
large enough that the effects of the interaction potential have died
off. If necessary, the solutions are transformed at $R_\mathrm{max}$
into a basis set in which $W_{aa'}$ and $\hat L^2$ are diagonal at
$R=\infty$ \cite{mlglez-mtnez:07a}, and the transformed channel
functions are matched to the standard scattering boundary conditions
\cite{Johnson:1973}. This gives the scattering matrix $S$, from which
all quantities of interest, such as state-to-state cross sections and
scattering lengths, may be calculated. Numerical details are given in
Sec.~\ref{sec2:xsections}.

\subsection{Basis set and matrix elements}
\label{sec2:basis+aHa'} We use a fully uncoupled basis set $\left| a
\right> \equiv \left|\alpha\right> \left|L M_L\right>$, where $\left|
\alpha \right> \equiv \left|i_\mathrm{N} m_{i\mathrm{N}}\right>
\left|i_\mathrm{H} m_{i\mathrm{H}}\right> \left|s m_s\right> \left|n
m_n\right>$ describes the state of the monomers, and $m_A$ (or $M_A$)
denotes the projection on the field axis of the vector operator
$\hat{A}$. None of the interactions considered here change the
electronic or nuclear spins, so we omit the labels $s$ and $i$ and
label our basis functions $(m_{i\mathrm{N}}, m_{i\mathrm{H}}, m_s, n,
m_n, L, M_L)$. A static magnetic field conserves both the projection
$M_{\rm tot}$ of the total angular momentum $F$ and the total parity
$P$ of the system. These are explicitly $M_{\rm tot} = m_{i\mathrm{N}}
+ m_{i\mathrm{H}} + m_s + m_n + M_L$ and $P = p_1 p_2 (-1)^L$, with
$p_1 = 1$ the parity of Mg($^1$S) and $p_2 = (-1)^{n+1}$ the parity of
NH($^3\Sigma^-$).  The matrix elements for the centrifugal, rotational,
electron spin-spin and interaction potential are diagonal in, and
independent of, the nuclear spin quantum numbers. The corresponding
expressions in our basis set are thus readily obtained from
Ref.~\cite{mlglez-mtnez:07a}.

\begin{widetext}
The Zeeman matrix elements are completely diagonal in the uncoupled
basis set,
 \begin{equation}
  \left<s m_s\right|
  \left<i_\mathrm{H} m_{i\mathrm{H}}\right|
  \left<i_\mathrm{N} m_{i\mathrm{N}}\right|
   \hat{\mathcal{H}}_\mathrm{Z}
  \left|i_\mathrm{N} m_{i\mathrm{N}}\right>
  \left|i_\mathrm{H} m_{i\mathrm{H}}\right>
  \left|s m_s\right> = B \left[g_S \mu_\mathrm{B} m_s
            - \mu_\mathrm{N} \left( g_\mathrm{N} m_{i\mathrm{N}}
                                 + g_\mathrm{H} m_{i\mathrm{H}} \right) \right].
  \label{eq:aHZa'}
 \end{equation}
Here and throughout this section, the matrix elements are fully
diagonal with respect to quantum numbers that do not appear explicitly
in the expression.

The Fermi contact term and the electron and nuclear spin-rotation terms
all share a similar structure, $\hat{\mathcal{H}}_\mathrm{j_1j_2} =
\kappa \hat{\jmath}_1 \cdot \hat{\jmath}_2$, where $\kappa$ is a scalar
while $\hat{\jmath}_1$ and $\hat{\jmath}_2$ are vector operators.  In
general, their matrix elements in a decoupled basis set, $\left|j_1
m_{j1}\right> \left|j_2 m_{j2}\right>$, are
\begin{eqnarray}
 &&\hspace{-25mm}\left<j_2 m_{j2}\right| \left<j_1 m_{j1}\right|
  \hat{\mathcal{H}}_\mathrm{j_1j_2}
 \left|j_1 m'_{j1}\right> \left|j_2 m'_{j2}\right> =
  \delta_{m_{j1} m'_{j1}} \delta_{m_{j2} m'_{j2}} \kappa\, m_{j1} m_{j2}
  \nonumber\\ && +
  \delta_{m_{j1} m'_{j1} \pm 1} \delta_{m_{j2} m'_{j2} \mp 1} \frac{\kappa}{2}
  \left[j_1(j_1 + 1) - m_{j1} m'_{j1} \right]^{1/2}
  \left[j_2(j_2 + 1) - m_{j2} m'_{j2} \right]^{1/2}.
 \label{eq:aHj1j2a'}
\end{eqnarray}
Such terms can mix functions with adjacent values of the projections of
$\hat{\jmath}_1$ and $\hat{\jmath}_2$, but preserve the sum $m_{12} =
m_{j1} + m_{j2}$.

The electron-nuclear spin dipolar interaction,
Eq.~\eqref{eq:Hsi_approx}, has matrix elements
 \begin{eqnarray}
  &&\hspace{-22mm}\left<n m_n\right| \left<s m_s\right| \left<i m_i\right|
   \hat{\mathcal{H}}_\mathrm{si}
   \left|i m'_i\right> \left|s m'_s\right> \left|n' m'_n\right> \nonumber\\
   &=& t_0 \sqrt{30}
   (-1)^{i-m_i+s-m_s-m_n}
   \left[i(i+1)(2i+1) s(s+1)(2s+1) (2n+1)(2n'+1)\right]^{1/2} \nonumber\\
   && \times
   \left(\begin{array}{ccc} n & 2 & n' \\
                            0 & 0 & 0  \end{array}\right)
   \sum_{q_1,q_2}
    \left(\begin{array}{ccc} 1   & 1   & 2  \\
                             q_1 & q_2 & -q \end{array}\right)
    \left(\begin{array}{ccc} i   & 1 & i    \\
                            -m_i & q_1 & m'_i \end{array}\right)
    \left(\begin{array}{ccc} s   & 1   & s    \\
                            -m_s & q_2 & m_s' \end{array}\right)
    \left(\begin{array}{ccc} n   & 2  & n'   \\
                            -m_n & -q & m'_n \end{array}\right),
  \label{eq:aHsia'}
 \end{eqnarray}
where $q \equiv q_1 + q_2$ and $\left(:::\right)$ is a 3-$j$ symbol.
This term produces couplings off-diagonal in one nuclear spin
projection $m_i$, along with $m_s$ and $m_n$ (keeping their sum
unchanged), with $\Delta n \equiv n' - n = 0$ or $\pm 2$. This latter
selection rule is required to conserve $p_2$.

Finally, the quadrupole interaction for the $^{14}$N nucleus has matrix
elements
 \begin{eqnarray}
  \left<n m_n\right|
  \left<i_\mathrm{N} m_{i\mathrm{N}}\right|
   \hat{\mathcal{H}}_\mathrm{Q,N}
  \left|i_\mathrm{N} m'_{i\mathrm{N}}\right>
  \left|n' m'_n\right>
  &=& \frac{(eq_0Q)_\mathrm{N}}{4} (-1)^{i_\mathrm{N} - m_{i\mathrm{N}} - m_n}
      \left[(2n+1)(2n'+1)\right]^{1/2}
      \left(\begin{array}{ccc} n & 2 & n' \\
                               0 & 0 & 0  \end{array}\right)
      \left(\begin{array}{ccc} i_\mathrm{N} & 2 & i_\mathrm{N} \\
                        -i_\mathrm{N} & 0 & i_\mathrm{N} \end{array}\right)^{-1}
   \nonumber\\ && \times
   \sum_p (-1)^p
   \left(\begin{array}{ccc} n    & 2 & n'   \\
                            -m_n & p & m'_n \end{array}\right)
   \left(\begin{array}{ccc} i_\mathrm{N} & 2 & i_\mathrm{N}    \\
                    -m_{i\mathrm{N}} & -p & m'_{i\mathrm{N}} \end{array}\right).
  \label{eq:aHQa'}
 \end{eqnarray}
These couple functions $\Delta n = 0, \pm 2$ and $\Delta
m_{i\mathrm{N}} = -\Delta m_n = 0, \pm 1, \pm 2$, thus preserving the
sum $m_{i\mathrm{N}} + m_n$ as well as $p_2$.
\end{widetext}

The first 3-$j$ symbol in Eqs.~\eqref{eq:aHsia'} and \eqref{eq:aHQa'}
implies that the electron-nuclear spin and quadrupolar interactions
have no direct off-diagonal matrix elements between $n = 0$ functions,
so that their dominant contribution for $n=0$ is a second-order effect
necessarily involving the $n = 2$ excited rotational state; this is
also the case for the electron spin-spin interaction.

\section{Results and Discussion}
\label{sec:results}

\begin{figure}[tbp]
\includegraphics[width=85mm]{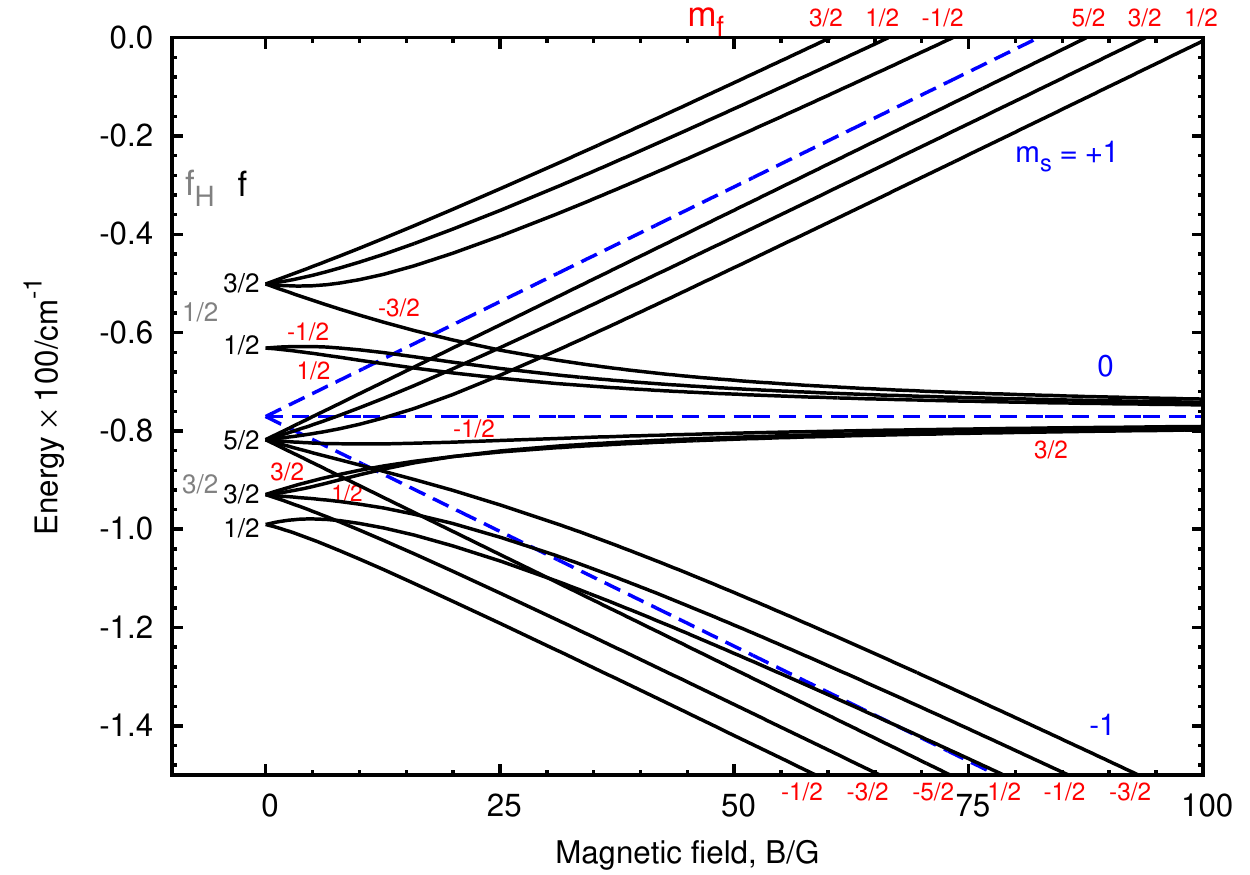}
\caption{(Color online) Magnetic field dependence of the energy
levels of $^{14}$NH($^3\Sigma^-$) correlating with the ground
rotational state, including (black, solid) and excluding (blue,
dashed) hyperfine terms.} \label{fig:thresholds}
\end{figure}

\subsection{Hyperfine levels}
\label{sec2:thresholds} The energy levels correlating with the ground
rotational state of an isolated $^{14}$NH($^3\Sigma^-$) molecule, in an
external magnetic field, are shown in Fig.~\ref{fig:thresholds}. The
solid lines correspond to the inclusion of all hyperfine interactions
as described in Sec.~\ref{sec2:Heff}, while the dashed lines show the
energy levels obtained when hyperfine terms are neglected.

The hyperfine-free levels are labeled at zero field by the eigenvalues
of the angular momentum $\hat{\jmath} = \hat{n} + \hat{s}$, which has
only one allowed value, $j=1$, for $n=0$ and $s=1$. In the presence of
a magnetic field, a level with quantum number $j$ splits into $2j + 1$
Zeeman components characterized by the projection $m_j$ onto the field
axis, and the corresponding eigenstates can be represented as
$\left|(ns)j m_j\right>$. For $n = 0$, $m_j = m_s$ and these states can
alternatively be labeled by $m_s$.

The pattern is much more complicated when hyperfine interactions are
included. In this case, the zero-field levels are labeled by the
eigenvalues of the total angular momentum $\hat{f}$ resulting from
coupling the nuclear spins $\hat{\imath}_\mathrm{N}$ and
$\hat{\imath}_\mathrm{H}$ with $\hat{n}$ and $\hat{s}$.  In general,
three or more angular momenta can be coupled using a variety of
schemes.  We start, as before, by first coupling $\hat{n}$ and
$\hat{s}$ to form $\hat{\jmath}$.  Then, given the molecular constants
in Table~\ref{tab:NHparameters}, it is convenient to couple
$\hat{\jmath}$ first with $\hat{\imath}_\mathrm{H}$ to produce a
resultant $\hat{f}_\mathrm{H}$.  Finally, $\hat{f}_\mathrm{H}$ is
coupled with $\hat{\imath}_\mathrm{N}$ to give $\hat{f}$. In the
particular case of $n = 0$, $j = s = 1$ and $f_\mathrm{H} =$~1/2, 3/2,
which produces levels with $f=1/2$ and 3/2 for $f_\mathrm{H} = 1/2$ and
$f=1/2$, 3/2 and 5/2 for $f_\mathrm{H} = 3/2$, as shown in
Fig.~\ref{fig:thresholds}. For $n=0$, the Fermi contact terms make by
far the largest contributions to the splittings, and levels with
$f_\mathrm{H} = 3/2$ lies below those with $f_\mathrm{H} = 1/2$ because
$b_\mathrm{F,H} < 0$.  In a magnetic field, each $f$ state further
splits into $2f + 1$ sublevels, producing a total of
$(2i_\mathrm{N}+1)(2i_\mathrm{H}+1)(2s+1) = 18$ components correlating
with the ground rotational state.  At low fields, below about 10~G, the
eigenfunctions are approximately represented as $\left|(ns)j,
(ji_\mathrm{H})f_\mathrm{H}, (f_\mathrm{H} i_\mathrm{N})f m_f \right>$.
In what follows, even though $f_\mathrm{H}$ and $f$ are not good
quantum numbers in a field, we use quantum numbers $(f_\mathrm{H}, f,
m_f)$ in parentheses to identify the states at low field. In addition,
levels will be labeled $\beta_i$ ($i = \overline{1,18}$) in order of
increasing energy at fields above 50~G, where $m_s$ is a nearly good
quantum number. With this convention, $\beta_1$--$\beta_6$ correspond
to $m_s = -1$, $\beta_7$--$\beta_{12}$ to $m_s = 0$, and
$\beta_{13}$--$\beta_{18}$ to $m_s = +1$.

For isolated NH, the total projection $m_f$ is a good quantum number.
Hence, as a function of field, states corresponding to different $m_f$
can cross while states of the same $m_f$ cannot.  However, in our model
Hamiltonian, states $\beta_{10}$ and $\beta_{13}$ (both with $m_f =
1/2$) are seen to cross at about 25~G.  This is a nonphysical effect
which results from neglecting the interaction between the nuclear spins
of N and H, usually written in the form $c_4 \hat{\imath}_\mathrm{N}
\cdot \hat{\imath}_\mathrm{H}$. For NH, $c_4$ is extremely small and
has not been measured experimentally; there is in reality an avoided
crossing between states $\beta_{10}$ and $\beta_{13}$, but it is
extremely tight.

At high field, the terms in the monomer Hamiltonian that mix states
with different values of $m_s$ are small with respect to the electron
Zeeman splitting and $m_s$ is well defined.  For $n=0$ at fields over
75~G, three groups of NH levels corresponding to $m_s = -1, 0$ and +1
can be identified, containing six hyperfine levels each.  In this
regime, the individual nuclear spin projections $m_{i\mathrm{N}}$ and
$m_{i\mathrm{H}}$ are also nearly conserved and the eigenfunctions are
well represented by individual basis functions $\left|\alpha\right>$.
Above 75~G, more than 93.5\% of any eigenstate is represented by a
single basis function $\left|\alpha\right>$. In the high-field limit we
label states with quantum numbers in square brackets, $[m_s,m_{i{\rm
N}},m_{i{\rm H}}]$.

All $m_s = +1$ states are low-field-seeking and therefore trappable in
a static magnetic trap. When hyperfine interactions are included, there
are six such levels with different values of $m_f$. In contrast, when
hyperfine terms are neglected, there is only one such state.

One state of particular interest is the \emph{spin-stretched} state, in
which $f$ and $m_f$ take their highest possible values. Except for
terms off-diagonal in $n$, this state is exactly represented by a
single basis function in either possible basis set, $(f_{\rm
H},f,m_f)=(3/2,5/2,+5/2)$ or $[m_s,m_{i{\rm N}},m_{i{\rm H}}] =
[+1,+1,+1/2]$. For $^{14}$NH, the spin-stretched state is $\beta_{15}$,
and lies below three levels from the $(f_\mathrm{H}, f) = (1/2, 3/2)$
manifold, $\beta_{16}$--$\beta_{18}$.

\subsection{Scattering cross sections}
\label{sec2:xsections} We have carried out scattering calculations
using the MOLSCAT package \cite{jmhutson:molscat14}, as modified to
handle collisions in external fields \cite{mlglez-mtnez:07a}.  The
coupled equations were propagated with the hybrid log-derivative Airy
method of Alexander and Manolopoulos \cite{mhalexander:87}, using a
fixed-step-size log-derivative propagator for $2.5 \le R \le 50$~\AA,
with $\Delta R = 0.025$~\AA, and a variable-step-size Airy propagator
for $50 \le R \le 250$~\AA.  To a good approximation, the computer time
is dominated by operations on relatively large matrices that scale with
the total number of channels $N$ as $\propto N^3$. The time needed to
perform a calculation including hyperfine interactions is thus
approximately $\left[ (2i_\mathrm{N}+1) (2i_\mathrm{H}+1) \right]^3 =
216$ times larger than that required for an equivalent calculation
neglecting them.  In order to make our calculations tractable, the
basis set used in the present work was reduced slightly from that used
in Ref.~\cite{aogwallis:09b}, to $n_\mathrm{max} = 5$ and
$L_\mathrm{max} = 6$.  Under these conditions $N \approx 1,500$, with
the actual number depending on the initial state, $M_{\rm tot}$ and
$P$.  The reduction in $n_\mathrm{max}$ and $L_\mathrm{max}$ does not
change the results by more than about 5\%.

NH molecules that undergo transitions between hyperfine levels of the
$m_s = +1$ manifold remain in a magnetically trappable state.  However,
the associated kinetic energy release ranges from 0.7 to 5.8~mK and
will either heat the trapped gas or eject one or both collision
partners from the trap. We will thus assume that all inelastic
processes have a negative impact on the success of sympathetic cooling.
In any case, as will be shown in Sec.~\ref{par:sigmaL0_bb'},
transitions with $\Delta m_s=0$ do not contribute appreciably to
inelasticity.

\subsubsection{General considerations}
\label{sec3:general}
\paragraph{Analytical model.}
\label{par:sigmaEB}

The total inelastic cross section may be decomposed into partial-wave
contributions,
\begin{equation}
\sigma_{\beta,\mathrm{inel}} = \sum_{\beta'\ne\beta, L,L'}
\sigma_{\beta L \rightarrow \beta' L'}.
\end{equation}
When inelastic scattering is weak compared to elastic scattering, the
first-order distorted-wave Born approximation \cite{mschild:book74}
provides relatively simple expressions for the off-diagonal $S$-matrix
elements.  Volpi and Bohn \cite{avolpi:02a} gave an analytical formula
for the threshold behavior of the partial inelastic cross sections
under these conditions,
\begin{equation}
 \sigma_{\beta L \rightarrow \beta' L'}(E,B) = \sigma^{L L'}_{\beta \beta'}
  E^{L-\frac{1}{2}}
  \left[ E + \Delta E_{\beta \beta'}(B) \right]^{L'+\frac{1}{2}}.
 \label{eq:sigmaEB}
\end{equation}
Here, $\sigma^{L L'}_{\beta \beta'}$ is a factor independent of the
collision energy $E$, while $\Delta E_{\beta \beta'}$ is the kinetic
energy released in the transition from the state $\beta$ to $\beta'$.
This formula was used in interpreting the energy and magnetic field
dependence of hyperfine-free scattering cross sections in Mg + NH
\cite{aogwallis:09b}.

When hyperfine terms are neglected, $\Delta E_{\beta\beta'}$ in
Eq.~\eqref{eq:sigmaEB} is simply $-g_S \mu_{\rm B} B \Delta m_s$ and
$\sigma^{L L'}_{\beta \beta'}$ is independent of the magnetic field
strength $B$. However, when hyperfine terms are included, the
$B$-dependence of $\Delta E_{\beta\beta'}$ is much more intricate, as
seen in Fig.~\ref{fig:thresholds}. In particular, for many transitions,
$\Delta E_{\beta\beta'}$ does \emph{not} approach zero at low magnetic
fields. This has substantial effects below about 50~G. At higher
fields, $\Delta E_{\beta\beta'}$ approaches its hyperfine-free form
even when hyperfine interactions are included. In addition, $\sigma^{L
L'}_{\beta \beta'}$ varies with $B$ because the character of the NH
eigenstates depends on the magnetic field.

\paragraph{Relaxation mechanisms.}
\label{par:relaxation} The quantum-mechanical theory of electron spin
relaxation in collisions of $^3\Sigma$ molecules with structureless
atoms, neglecting hyperfine effects, was developed by Krems and
Dalgarno \cite{rvkrems:04a}. In general, inelastic collisions are
driven by the anisotropy of the interaction potential, but this does
not have matrix elements that are off-diagonal in $m_s$. Collisions
that change $m_s$ thus occur only because it is not strictly a good
quantum number, and basis functions with different values of $m_s$ are
mixed by terms in the monomer Hamiltonian. In the absence of hyperfine
interactions, the only such terms are the spin-spin and spin-rotation
Hamiltonians.

In the NH($^3\Sigma^-$) case that we study here, $\lambda_\mathrm{ss}
\gg \gamma$ and the electron spin-spin terms are dominant. There are no
matrix elements of $\hat{\cal H}_{\rm ss}$ between $n=0$ states,
because $m_n$ cannot change from 0. In the absence of hyperfine
interactions, the effect of $\hat{\cal H}_{\rm ss}$ is to mix into the
ground state $(m_s, n, m_n, M_L) = (m_s, 0, 0, M_L)$ a small amount of
rotationally excited functions $(m_s-q, 2, q, M_L)$ with
$q=0,\pm1,\pm2$. The potential anisotropy (principally $V_2$) can then
drive transitions between these mixed states. Spin relaxation for $n=0$
states thus proceeds mainly via the combination of the spin-spin
interaction and the potential anisotropy, and leads to transitions with
$\Delta m_s$ compensated by a change in $M_L$ to conserve $M_{\rm
tot}$. If $L=0$, this requires $L'>0$ and the corresponding cross
sections are therefore \emph{centrifugally suppressed} at low energies
by barriers in the outgoing channels. In what follows, we will refer to
this as the \emph{main} relaxation mechanism.

When hyperfine coupling is included, states with different values of
$m_s$ and $m_i$ are strongly mixed at low fields. For the $n=0$ states,
this mixing is almost entirely due to the Fermi contact interactions.
Under these circumstances the main mechanism can drive all possible
transitions among the $n=0$ states, although with varying degrees of
centrifugal suppression as described below. However at high fields,
once the Zeeman splittings are large compared to the Fermi contact
interactions, $m_s$ and $m_i$ become nearly good quantum numbers. Since
the main mechanism does not affect the nuclear spin projections, there
is a propensity for transitions with $\Delta m_i=0$.

\subsubsection{Dependence on collision energy}
\label{sec3:sigmaE}
\begin{figure}[t]
\includegraphics[width=85mm]{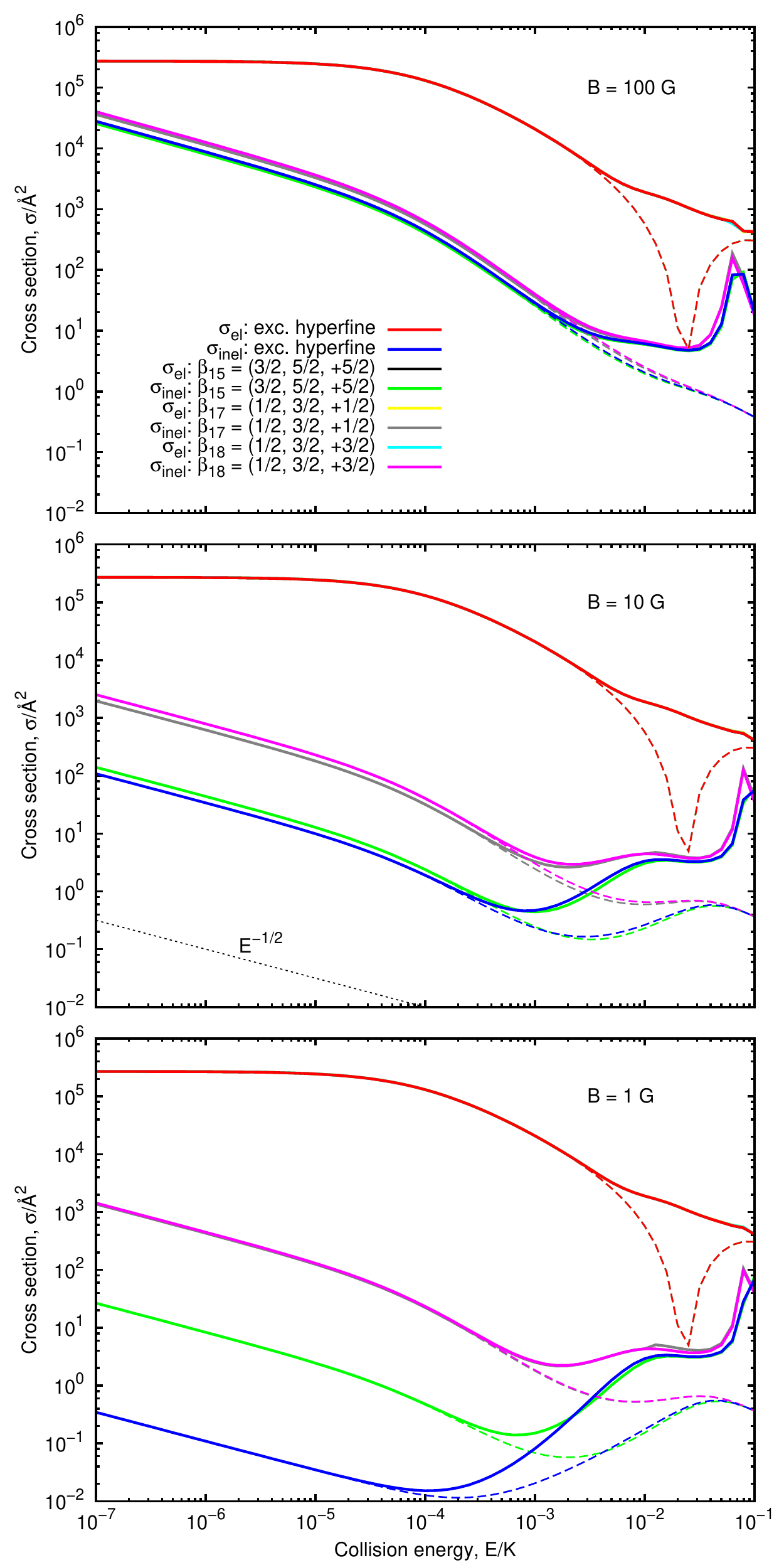}
\caption{(Color online) Elastic and total inelastic cross sections as a
function of collision energy, for various magnetic fields and initial
states.  Solid lines include $s$, $p$ and $d$-wave contributions, while
dashed lines are $s$-wave cross sections.  Dotted lines show power-law
to guide the eye.}
 \label{fig:xsecE_B1+10+100G}
\end{figure}
Fig.~\ref{fig:xsecE_B1+10+100G} shows the elastic and total inelastic
cross sections as a function of energy at magnetic fields $B = 1$, 10
and 100~G, for collisions starting in the spin-stretched state,
$\beta_{15} = (f_{\rm H},f,m_f) = (3/2, 5/2, +5/2)$, and the two
highest-lying states, $\beta_{17} = (1/2, 3/2, +1/2)$ and $\beta_{18} =
(1/2, 3/2, +3/2)$. At very low energies, the scattering of incoming
partial waves with $L \ne 0$ is suppressed by centrifugal barriers; our
calculations include all contributions from incoming $s$, $p$ and $d$
waves ($L=0$ to 2), which gives integral cross sections that are
converged at collision energies up to about 40~mK. The dashed lines in
Fig.~\ref{fig:xsecE_B1+10+100G} show pure s-wave cross sections, and it
may be seen that collisions with $L>0$ become significant above $E
\approx 10^{-4}$~K.

The dependence on the magnetic field will be analyzed in detail in the
next section. However, in general terms it is clear that at high fields
(100~G and above), the cross sections including hyperfine interactions
are quite similar to those from hyperfine-free calculations, while at
lower fields they are very different. In particular, the suppression of
inelastic cross sections that occurs in hyperfine-free calculations at
low fields and low energies is much reduced (resulting in larger
inelastic cross sections) when hyperfine interactions are included.

The major effect of hyperfine interactions is that, at low fields, they
increase the kinetic energy release and thus reduce the centrifugal
suppression of the inelastic scattering. At 1~G, for example, the
$\beta_{17}$ and $\beta_{18}$ states can relax to channels with kinetic
energy releases of up to 7.2~mK, while the corresponding value for the
spin-stretched $\beta_{15}$ state is only 2.6~mK. When hyperfine
coupling is excluded, however, the kinetic energy release at 1~G is
only about 270~$\mu$K.

For simplicity, let us consider the case of $s$-wave scattering. The
projection of the total angular momentum, $M_{\rm tot} = m_f + M_L$
(with $m_f = m_s + m_{i\mathrm{N}} + m_{i\mathrm{H}}$) is conserved in
a collision, so $L' \ge |M'_L| = |\Delta m_f|$.  In addition,
conservation of parity requires $L'$ to be even. For scattering from
the spin-stretched state, $\beta_{15}$, the main relaxation channels
have $m'_f = 1/2$ and $3/2$, for which $L'$ must be at least 2.
Channels with $L' = 2$ dominate at low energies, because $d$-wave
centrifugal barriers (height 23~mK) are much lower than $g$-wave
barriers (height 140~mK). Transitions to levels for which $L'=2$ is not
possible, such as $(m'_f, L'_\mathrm{min}) = (-1/2, 4)$, $(-3/2, 4)$,
$(-5/2, 6)$ have negligible contributions at the magnetic fields
considered here.

For molecules that are initially in a non-spin-stretched state,
inelastic collisions with $\Delta m_f = 0$ are possible. There are then
relaxation channels with $L' = 0$, which are \emph{not} centrifugally
suppressed. However, transitions to these states are made possible only
by invoking hyperfine couplings. Such transitions make very little
contribution to the total inelastic cross section, which remains
dominated by the (centrifugally suppressed) main mechanism. Even for
molecules in non-spin-stretched states, the main effect of hyperfine
interactions is through an increased kinetic energy release that helps
overcome the centrifugal barriers at low fields and low energies.

In general, the elastic scattering depends on the phases of diagonal
elements of the $S$-matrix, which are only very slightly affected by
the inclusion of hyperfine terms. Therefore, as seen in
Fig.~\ref{fig:xsecE_B1+10+100G}, the elastic cross sections including
hyperfine interactions are very similar to the hyperfine-free results
at all energies and fields.

\subsubsection{Dependence on magnetic field}
\label{sec3:sigmaB}
\paragraph{State-to-state $s$-wave cross sections.}
\label{par:sigmaL0_bb'}
\begin{figure}[t]
\includegraphics[width=85mm]{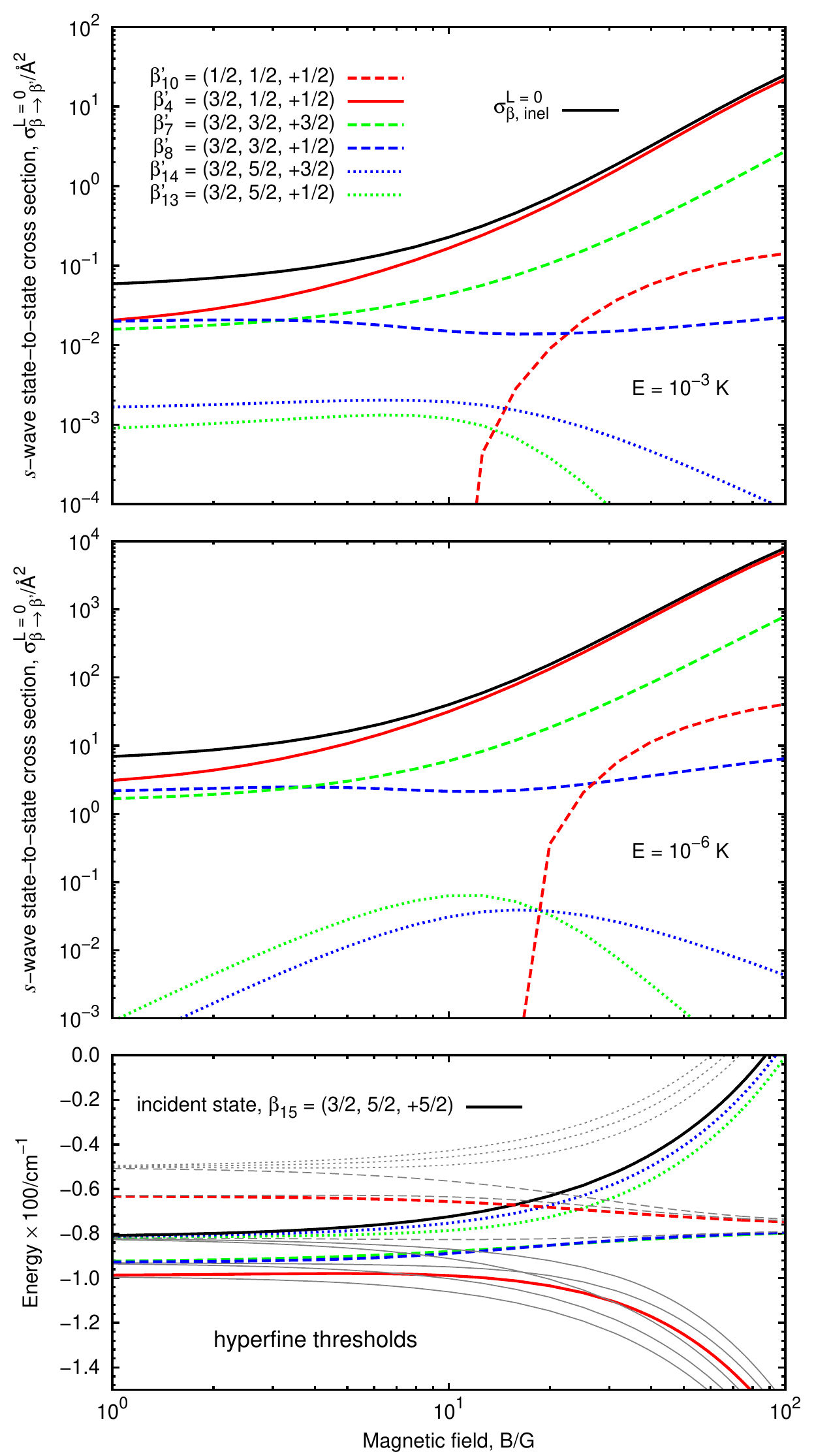}
\caption{(Color online). State-to-state $s$-wave inelastic cross
sections for collisions originating in the spin-stretched state
$\beta_{15}$, as a function of magnetic field, for collision energies
of $10^{-3}$~K (top panel) and $10^{-6}$~K (center panel). The bottom
panel shows the initial and final states energies, color-coded as for
the cross sections. Solid, dashed and dotted lines represent states
with $m_s = -1$, 0 and $+1$ respectively.}
\label{fig:xsecB_th2.5_l0_s2s}
\end{figure}

Fig.~\ref{fig:xsecB_th2.5_l0_s2s} shows the state-to-state $s$-wave
inelastic cross sections $\sigma^{L = 0}_{\beta \rightarrow \beta'}
\equiv \sum_{L'} \sigma_{0\beta \rightarrow L'\beta'}$ as a function of
magnetic field $B$, for collisions starting in the spin-stretched
state, $\beta_{15}$, with quantum numbers $(f_{\rm
H},f,m_f)=(3/2,5/2,+5/2)$ and $[m_s,m_{i{\rm N}},m_{i{\rm
H}}]=[+1,+1,+1/2]$. There are six main contributions, all to channels
with $L' = 2$, corresponding to $\beta'_4$ ($\Delta m_s = -2$),
$\beta'_7$, $\beta'_8$ and $\beta'_{10}$ ($\Delta m_s = -1$), and
$\beta'_{13}$ and $\beta'_{14}$ ($\Delta m_s = 0$).  The corresponding
final-state energies are shown color-coded in the bottom panel.

At low fields, where the main mechanism can drive all possible
transitions, the state-to-state cross sections are governed by the
kinetic energy release. The largest cross section at $E=10^{-6}$~K is
to $\beta'_4=(3/2,1/2,+1/2)$, closely followed by $\beta'_7$ and
$\beta'_8$, which are $(3/2,3/2,+3/2)$ and $(3/2,3/2,+1/2)$
respectively. These channels have the largest kinetic energy release
and therefore experience less centrifugal suppression. The relatively
minor channels $\beta'_{13}$ and $\beta'_{14}$ have zero kinetic energy
release at low field, while $\beta'_{10}$ is energetically accessible
only at fields above about 15~G (slightly dependent on $E$), as seen in
the bottom panel.

At high fields, where $m_s$ and $m_i$ become nearly good quantum
numbers, transitions with $\Delta m_i=0$ are favored. The two strongest
channels are $\beta'_4=[-1,+1,+1/2]$ ($\sim$90\%) and
$\beta'_7=[0,+1,+1/2]$ ($\sim$10\%). The former is stronger because of
the larger kinetic energy release associated with $\Delta m_s=-2$. The
largest cross sections to channels with $\Delta m_i\ne0$ are those to
$\beta'_{10}=[0,+1,-1/2]$ and $\beta'_8=[0,0,+1/2]$, with the former
making a greater contribution because the Fermi contact interaction is
stronger for H than for N. Transitions to $\beta'_{13}=[+1, -1, +1/2]$
and $\beta'_{14}=[+1, 0, +1/2]$ are weak both because of the change in
$m_{i,{\rm N}}$ and because they do not change $m_s$ and thus have a
small kinetic energy release.

The relative state-to-state cross sections are fairly insensitive to
the collision energy, as seen by comparing the top and middle panels in
Fig.~\ref{fig:xsecB_th2.5_l0_s2s}. The only qualitative difference is
in the cross sections to $\beta'_{13}$ and $\beta'_{14}$, for which the
kinetic energy release is zero at zero field. The inelasticity to these
states is the most affected by centrifugal suppression. The outgoing
kinetic energy for these channels is thus dominated by $E$, and the
increase with collision energy at fields below $\sim$5~G is simply due
to a larger probability of tunneling through the outgoing centrifugal
barriers.

\begin{figure}[!t]
\includegraphics[width=85mm]{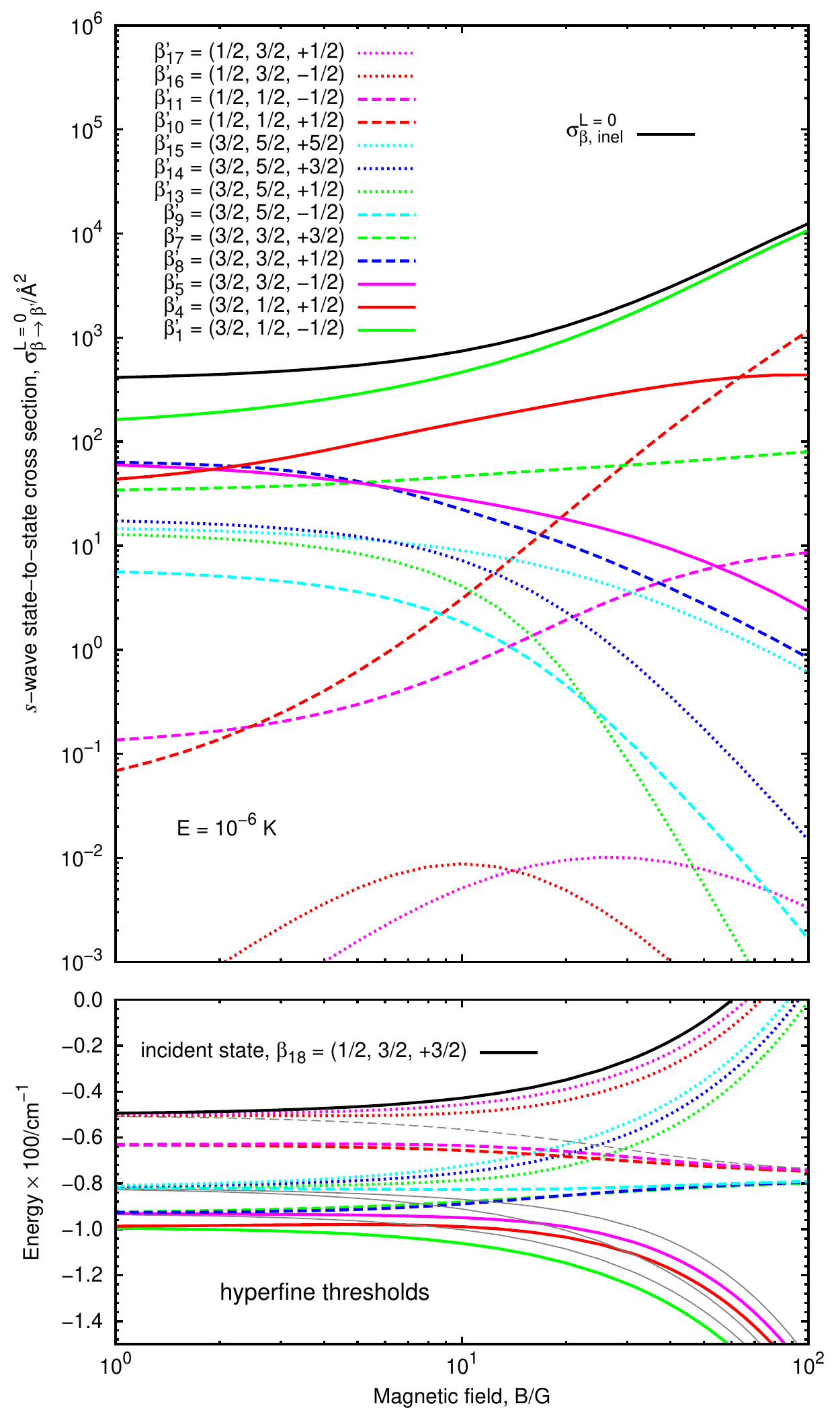}
\caption{(Color online). State-to-state $s$-wave inelastic cross
sections for collisions originating in the uppermost hyperfine state,
$\beta_{18}$, as a function of magnetic field, for a collision energy
of $10^{-6}$~K (upper panel). The lower panel shows the initial and
final states energies, color-coded as for the cross sections. Solid,
dashed and dotted lines represent states with $m_s = -1$, 0 and $+1$
respectively.} \label{fig:xsecB_th1.5t_l0_s2s}
\end{figure}

The behavior of the state-to-state cross sections from other states in
the $m_s = +1$ manifold is considerably more complicated.
Fig.~\ref{fig:xsecB_th1.5t_l0_s2s} shows the state-to-state cross
sections for collisions that start in $\beta_{18}$, which at low field
is $(1/2,3/2,+3/2)$ and at high field is $[+1, +1, -1/2]$. Once again
the cross sections at low field are mostly governed by the kinetic
energy release.

At high field, the strongest transitions are those to
$\beta'_{1}=[-1,+1,-1/2]$ and $\beta'_{10}=[0,+1,-1/2]$, which are
driven by the main mechanism with no change in $m_i$ quantum numbers.
As before, the transition to $m_s=-1$ is stronger because of the larger
kinetic energy release. The next strongest are to
$\beta'_4=[-1,+1,+1/2]$ and $\beta'_7=[0,+1,+1/2]$, with $\Delta
m_{i,{\rm H}}=+1$.

As discussed above, for non-spin-stretched states such as $\beta_{18}$
it is possible to relax $m_s$ while conserving $m_f$. This is the case
for transitions to $\beta'_7$ and $\beta'_{14}$, which are dominated by
$L' = 0$ and therefore are not centrifugally suppressed. However, it is
clear from Fig.~\ref{fig:xsecB_th1.5t_l0_s2s} that the centrifugally
unsuppressed channels are not the dominant ones, even at very low
field: the hyperfine splittings release enough kinetic energy that the
main mechanism dominates over centrifugally unsuppressed transitions at
all values of $B$.

\begin{figure}[!t]
\includegraphics[width=85mm]{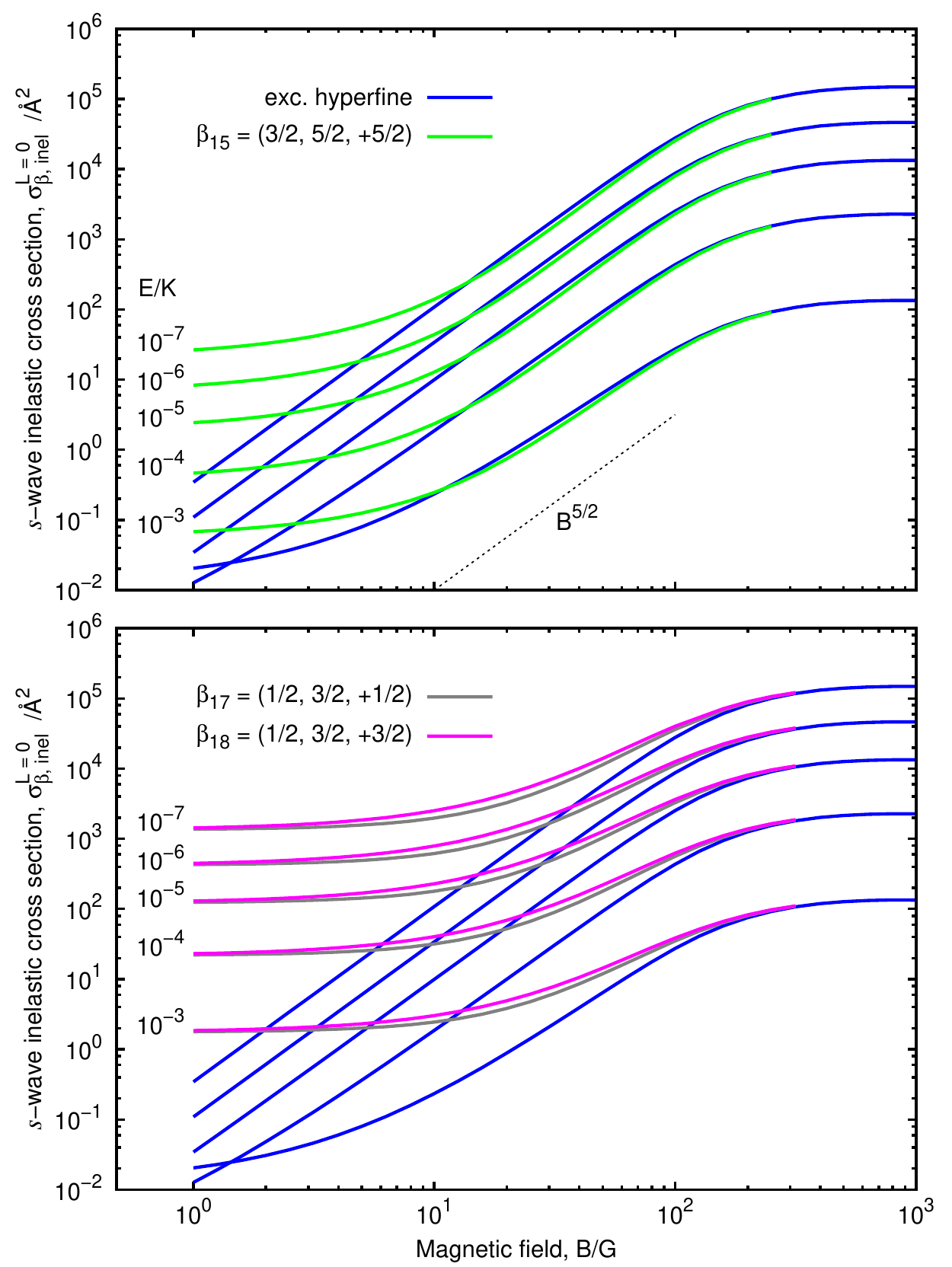}
\caption{(Color online). Total $s$-wave inelastic cross sections as a
function of magnetic field for a variety of collision energies of
$10^{-3}$~K (top panel) and $10^{-6}$~K (center panel). The
states are color-coded as in Fig.~\ref{fig:xsecE_B1+10+100G}. The
dotted line shows the $B^{5/2}$ behavior in regime (2) (see text).}
\label{fig:xsecB_l0}
\end{figure}

\paragraph{Total $s$-wave inelastic cross sections.}
\label{par:sigmaL0_b} The behavior of the total $s$-wave inelastic
cross sections with magnetic field, for the three initial states
studied above, is shown in Fig.~\ref{fig:xsecB_l0} for a range of
collision energies. If hyperfine interactions are neglected, the
quantity $\sigma^{0L'}_{\beta\beta'} \equiv \sigma^{0L'}_{m_s m'_s}$ of
Eq.\ \eqref{eq:sigmaEB} is independent of $B$, $\Delta E_{\beta\beta'}$
is given by $-g_S\mu_{\rm B}B\Delta m_s$, and three main regimes are
observed \cite{aogwallis:09b}: (1) at low enough fields, the inelastic
cross sections flatten out to a zero-field value proportional to $E^2$;
(2) as the field increases, $\sigma^{L=0}_\mathrm{inel}$ enters a
region of $B^{5/2}$ dependence, given by the increasing probability of
tunneling through the $d$-wave centrifugal barrier in the dominant
outgoing channel(s); and (3) at high enough fields (above about 100~G)
the $d$-wave centrifugal barriers are overcome and
$\sigma^{L=0}_\mathrm{inel}$ again approaches a field-independent
value, this time proportional to $E^{-1/2}$.

The inclusion of hyperfine terms modifies both the qualitative behavior
in regime (1) and the boundaries of regime (2). First, at very low
fields and collision energies, the state-to-state cross sections become
nearly constant at a field-free value that is much larger than when
hyperfine coupling is neglected. The kinetic energy release in this
region is dominated by $\Delta E_{\beta\beta'}$, which for some
outgoing channels does not approach zero as the field decreases. The
field-free cross section is proportional to $E^{-1/2}$ at the lowest
energies, though deviations from this occur at energies high enough
that the outgoing energy is no longer dominated by $\Delta
E_{\beta\beta'}$. Increasing the magnetic field alters $\Delta
E_{\beta\beta'}$, particularly above $\sim$10~G, but also changes
$\sigma^{0L'}_{\beta\beta'}$ because the spin character of the monomer
eigenfunctions changes. This leads to a non-power-law increase in the
cross sections up to the onset of regime (2). It may again be noted
that, although $L' = 0$ is possible for initial states other than the
spin-stretched state, the hyperfine splitting at zero field provides
enough kinetic energy release for the main mechanism to dominate spin
relaxation, even though it is suppressed by $d$-wave outgoing barriers.

\subsubsection{Prospects for sympathetic cooling}
\label{sec3:prospectsSC}
\begin{figure*}[!t]
 \includegraphics[width=170mm, trim=0mm 2mm 22mm 2mm]{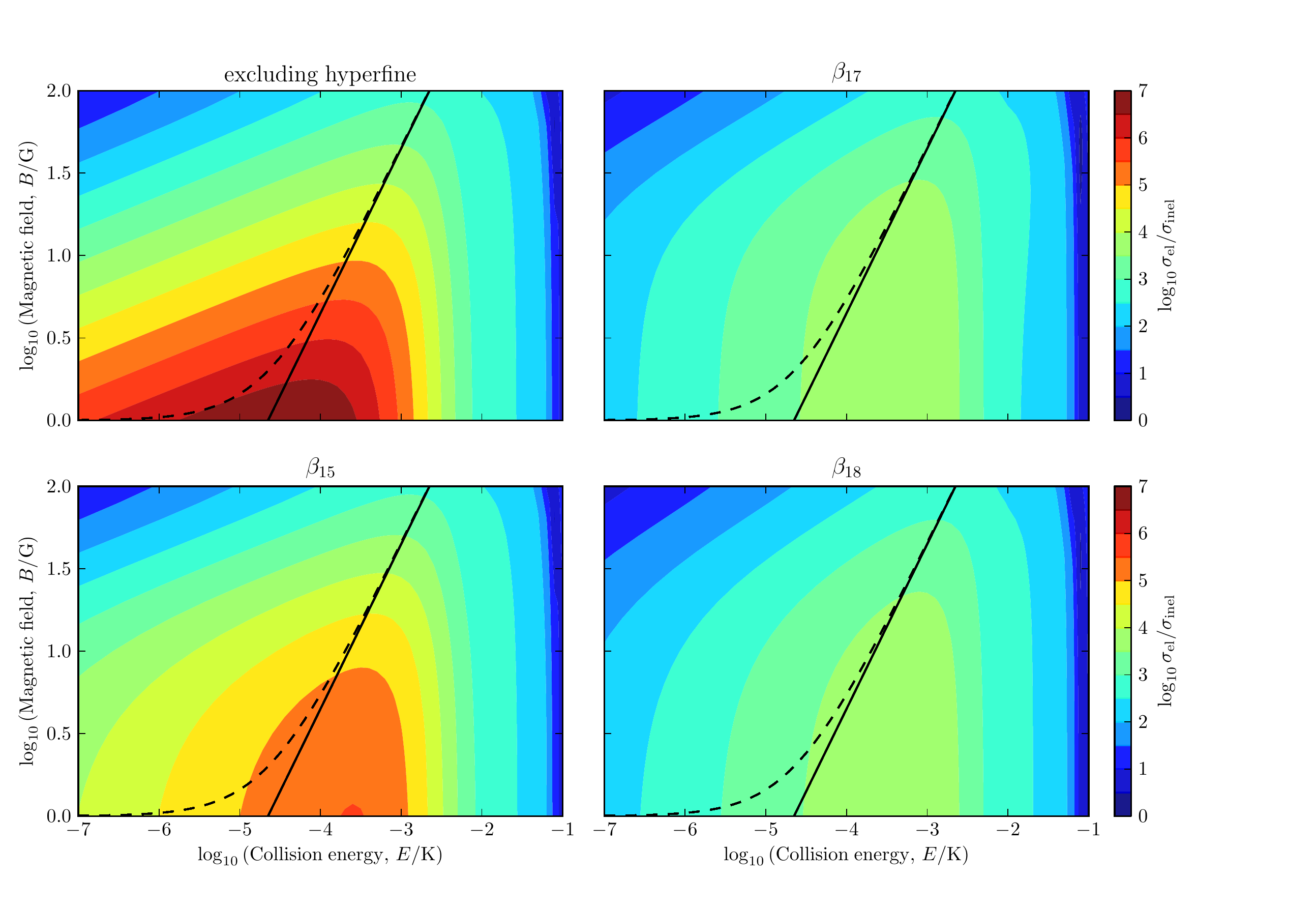}
 \caption{(Color online) Contour plots of the ratio of elastic to total
  inelastic cross section as a function of collision energy and magnetic field.
  The panels correspond to calculations excluding (top left) and including
  (rest) hyperfine terms.  The lines show the conditions sampled by 99.9\% of
  molecules trapped in the $m_s = +1$ state, in an unbiased trap (solid) and a
  trap with a bias field of $B = 1$~G (dashed).}
 \label{fig:xratioEB}
\end{figure*}

Trap losses in a static trap are fundamentally caused by four
phenomena: spin relaxation, background gas collisions, blackbody
radiation and non-adiabatic transitions to untrappable states.
Non-adiabatic transitions, which are one-body transitions that can
occur at points in the trap where different states are near-degenerate,
have important consequences for trap design. In particular, it is well
known for atomic systems that substantial losses can occur at the
center of magnetic quadrupole traps, where the magnetic field is zero
\cite{wpetrich:95}. Near this point, states with different values of
$m_f$ are degenerate and the trapping field varies very fast with
position, so that atoms can undergo non-adiabatic transitions (Majorana
flops \cite{emajorana:32}) when they pass close to the trap center. For
atoms, Majorana transitions can be effectively suppressed by applying a
small bias field (1~G or less) that removes the zero-field point.
Similar effects have been observed for molecules in electrostatic traps
\cite{Kirste:2009}, and can again be suppressed by applying a bias
field.

For molecules there are crossings that occur at non-zero field, as
shown in Fig.\ \ref{fig:thresholds}. Magnetically trapped NH molecules
in states $\beta_{13}$, $\beta_{14}$ and $\beta_{15}$ might conceivably
undergo transitions to untrapped states $\beta_{10}$, $\beta_{11}$ and
$\beta_{12}$ in the vicinity of crossings that occur between 15 and
30~G. However, away from the trap center the molecules experience a
field that varies only slowly as they move, and under such
circumstances the transition probabilities will be very low. We
therefore expect that a small bias field of 1~G or less will be
sufficient to suppress one-body losses for NH and other similar
molecules. Even if such losses do prove significant, the states
$\beta_{16}$, $\beta_{17}$ and $\beta_{18}$ are immune to them except
near a zero-field point.

The major loss mechanism in sympathetic cooling thus arises from
inelastic collisions. Provided that the absolute values of the elastic
cross sections are large enough to provide cooling before the molecules
are lost to black-body radiation or non-adiabatic transitions, the key
quantity is the {\em ratio} of elastic to total inelastic cross
sections, which must be greater than about 100 for sympathetic cooling
to proceed.

Both Mg-NH and NH-NH \cite{lmcjanssen:11a, lmcjanssen:11b} collisions
may cause transitions to untrapped states or release enough energy to
eject the molecules from the trap. Fig.~\ref{fig:xratioEB} shows
contour plots of the ratio of elastic to total inelastic cross section
as a function of $E$ and $B$ for Mg-NH. The top-left panel shows the
results when hyperfine terms are neglected, and the remainder show the
results when hyperfine terms are included, for the spin-stretched and
two highest-lying hyperfine states.

Trapped NH molecules in state $\beta$ at temperature $T$ will be
distributed according to a Boltzmann distribution with density $\rho$
given by
\begin{equation}
\rho/\rho_0 = \exp\left(\frac{E_\beta(0)-E_\beta(B)}{k_{\rm B}T}\right).
\end{equation}
At any given temperature on the energy axis of Fig.\
\ref{fig:xratioEB}, only about 0.1\% of molecules will experience
fields corresponding to energies greater than $6k_{\rm B}T$. The
diagonal lines in Fig. \ref{fig:xratioEB} show the maximum fields
sampled by over 99.9\% of the molecules trapped in one of the hyperfine
levels of the $m_s = +1$ manifold.  These correspond, respectively, to
a unbiased trap with zero magnetic field at the center (solid lines),
and a trap with a bias field of 1~G, to prevent Majorana transitions
(dashed lines). Pre-cooled molecules would enter the trap from the
right-hand side of the panels in Fig.~\ref{fig:xratioEB} and then move
to the left as they are cooled, remaining below the line appropriate
for the trap in use.

The ratio of elastic to inelastic cross sections exceeds 100 at
temperatures below 10~mK for all three hyperfine states, and is thus
favorable for sympathetic cooling to work, provided the molecules can
be precooled to this temperature. This agrees with the hyperfine-free
results of Wallis and Hutson \cite{aogwallis:09b}. At lower
temperatures, the ratios of elastic to inelastic collisions are not as
favorable as in hyperfine-free calculations, but are nevertheless
adequate to reach temperatures below 1 $\mu$K. Trapping molecules in
the fully spin-stretched state may be particularly advantageous,
especially for molecules with stronger hyperfine interactions than NH.

\section{Summary and Conclusions}
\label{sec:summary}

We have investigated the effect of hyperfine interactions on
spin-relaxation collisions of NH with Mg in the cold and ultracold
regimes. We find that hyperfine interactions make substantial changes
to inelastic collision rates at temperatures below about 10~mK and
magnetic fields below about 20~G. The major effect arises because
hyperfine interactions modify the kinetic energy released in
spin-relaxation collisions. When hyperfine interactions are neglected,
the kinetic energy decreases to zero as the field is decreased, but
when hyperfine interactions are included the kinetic energy release is
significant for most transitions even at zero field. For s-wave
collisions, the kinetic energy release helps overcome the $d$-wave
centrifugal barriers that suppress spin-relaxation collisions and thus
leads to larger inelastic cross sections.

Hyperfine interactions also introduce new mechanisms for
spin-relaxation collisions. For initial states that are not
spin-stretched, the cross sections for some of these are centrifugally
unsuppressed. However, for Mg-NH, where the hyperfine interactions are
quite weak, the centrifugally unsuppressed mechanisms make only a small
contribution to total inelastic cross sections at the collision
energies and fields studied here. It is nevertheless possible that
centrifugally unsuppressed channels may be important in other systems,
with either stronger hyperfine interactions or weaker competing spin
relaxation mechanisms.

The most important hyperfine effects for Mg-NH arise from the Fermi
contact interactions. These determine both the composition of the $n=0$
states in terms of spin functions and the low-field energy level
pattern (and hence the kinetic energy release). Other hyperfine terms
have only very small effects for transitions between $n=0$ states.
Indeed, we have repeated the calculations of the state-to-state cross
sections including {\em only} the Fermi contact interactions and obtain
almost identical results.

Our results for Mg-NH($^3\Sigma^-$) may be compared with those of
Tscherbul \emph{et al.}\ \cite{tvtscherbul:07} for He-YbF($^2\Sigma$).
For YbF the main mechanism of electron spin relaxation considered here,
driven by the electron spin-spin coupling, does not exist and is
replaced by a higher-order and much weaker mechanism driven by the
spin-rotation interaction. Under these circumstances, combined electron
and nuclear spin relaxation, driven by the electron-nuclear dipolar
interaction, is in relative terms much more important. However,
Tscherbul \emph{et al.}\ did not focus on the regime where hyperfine
energies make important contributions to the kinetic energy release.

We have considered the prospects for sympathetic cooling of NH by Mg,
which were previously explored in hyperfine-free calculations by Wallis
and Hutson \cite{aogwallis:09b}. We have calculated the ratio of the
elastic to inelastic cross section as a function of energy and magnetic
fields for several magnetically trappable hyperfine states of NH. Even
though hyperfine interactions increase inelastic cross sections at low
energies and magnetic fields, the ratio remains high enough for
sympathetic cooling to proceed if the NH molecules can be pre-cooled to
about 10~mK.

Molecular hyperfine interactions are also likely to be important in
developing techniques for controlling ultracold molecules. The
low-lying excited states afforded by hyperfine splittings can support
near-threshold levels that will produce magnetically tunable Feshbach
resonances. Once molecules such as NH have been cooled to the ultracold
regime, it will be possible to use such resonances both to control
collisions by adjusting the scattering length and to create polyatomic
molecules by magnetoassociation, as has already been achieved for
alkali-metal atoms \cite{jmhutson:06, tkohler:06}.

\begin{acknowledgments}
The authors are grateful to EPSRC for funding.
\end{acknowledgments}
\bibliographystyle{apsrev}
\bibliography{./all_refs,../../all}

\end{document}